\documentclass[conference,10pt]{IEEEtran}
\usepackage{balance} % For balanced columns on the last page
\usepackage{amssymb}
\usepackage{graphicx}
\usepackage{epstopdf}
\usepackage{enumerate}
\usepackage{multirow}
\usepackage{amsmath}

\usepackage{soul, color}
\usepackage{pseudocode}
\usepackage[noend]{algpseudocode}
\usepackage{algorithm}

\newcommand{\timediff}[1]{\mbox{time-diff}}
\newcommand*\Let[2]{\State #1 $\gets$ #2}

\begin{document}

\title{Temporal Phase Shifts in SCADA Networks}
%\titlenote{(Produces the permission block, and copyright information). For use with SIG-ALTERNATE.CLS. Supported by ACM.}
%\subtitle{[Extended Abstract]
%\titlenote{A full version of this paper is available as
%\textit{Author's Guide to Preparing ACM SIG Proceedings Using
%\LaTeX$2_\epsilon$\ and BibTeX} at
%\texttt{www.acm.org/eaddress.htm}}}

\author{\IEEEauthorblockN{Chen Markman\IEEEauthorrefmark{1},
			 		     Avishai Wool\IEEEauthorrefmark{2},
                         Alvaro A. Cardenas\IEEEauthorrefmark{3}
                    }
		
\IEEEauthorblockA{\IEEEauthorrefmark{1}\IEEEauthorrefmark{2} School of Electrical Engineering, Tel-Aviv University, Israel, {\tt chen.markman@gmail.com, yash@eng.tau.ac.il}\\
\IEEEauthorrefmark{3}{Department of Computer Science, University of Texas at Dallas, {\tt alvaro.cardenas@utdallas.edu}}
    }
}

\maketitle

\begin{abstract}
In Industrial Control Systems (ICS/SCADA), machine to machine data traffic is highly periodic. Previous work showed that in many cases, it is possible to create an automata-based model of the traffic between each individual Programmable Logic Controller (PLC) and the SCADA server, and to use the model to detect anomalies in the traffic. 
When testing the validity of previous models, we noticed that overall, the models have difficulty in dealing with communication patterns that change over time.
In this paper we show that in many cases the traffic exhibits phases in time, where each phase has a unique pattern, and the transition between the different phases is rather sharp. We suggest a method to automatically detect traffic phase shifts, and a new anomaly detection model that incorporates multiple phases of the traffic. Furthermore we present a new sampling mechanism for training set assembly, which enables the model to learn all phases during the training stage with lower complexity. The model presented has similar accuracy and much less permissiveness compared to the previous general DFA model. Moreover, the model can provide the operator with information about the state of the controlled process at any given time, as seen in the traffic phases.
\end{abstract}

%
%  Use this command to print the description
%
%%%\printccsdesc

\section{Introduction}

\subsection{Background}
Industrial Control Systems (ICS) are used for monitoring and controlling numerous industrial systems and processes such as chemical plants, electric power generation, transmission and distribution systems, oil and gas systems, water distribution networks, and waste-water treatment facilities. ICS is a general term that encompasses several types of control systems, including Distributed Control Systems (DCS), Supervisory Control And Data Acquisition (SCADA) systems, and other control system configurations~\cite{stouffer}. An automation system managed by a single vendor is usually referred to as a DCS, while SCADA systems usually refer to different stations distributed over large geographical areas.

%ICS are used in critical infrastructure assets such as  

ICS typically incorporate sensors and actuators that are controlled by Programmable Logic Controllers (PLCs), and which are themselves managed and monitored by a SCADA server or a Human Machine Interface (HMI). 

ICS have a strategic significance due to the potentially serious consequences a fault or malfunction can cause to our critical infrastructures. ICS were originally designed for serial communications, and were built on the premise that all the operating entities would be legitimate, properly installed, perform the intended logic, and follow the protocols of the system. As recent attacks have shown, it is no longer safe to assume that all devices in an ICS are trusted, therefore, deploying an anomaly detection system in an ICS network is an important defensive measure.

\subsection{Related work}

\subsubsection{Attacks}
The susceptibility of ICS systems to attacks has been known for more than a decade. Byres et al.~\cite{byres2004use} illustrated different attack trees on SCADA systems using the Modbus/TCP industrial network protocol. They found that compromising the slave (PLC) or the master (HMI) gives the attacker the highest impact on the SCADA system. For instance, an attacker that gains access to the  HMI could change set points of operation and other values in the PLCs.  Alternately, an attacker can perform a Man In The Middle attack between a PLC and HMI and ``feed'' the HMI misleading data, allegedly coming from the 
%exploited 
PLC.

More recently, at BlackHat USA 2015 Klick et al. \cite{klickinternet} demonstrated injection of malware into a SIMATIC S7-300 PLC without service disruption.
In a follow on work, \cite{spennebergplc} demonstrated the feasibility of a PLC worm.
The worm spreads internally from one PLC to 
%other target PLCs.
another. During the infection phase the worm scans the network for new targets (PLCs).

Digital attacks that cause physical destruction of equipment are also becoming more common. Perhaps most notably is the attack on an Iranian nuclear facility in 2010 (Stuxnet) to sabotage centrifuges at a uranium enrichment plant. The Stuxnet malware \cite{falliere2011w32,langner2011stuxnet} implemented an attack by changing centrifuge operating parameters in a pattern that damaged the equipment---while sending normal status messages to the HMI to hide the fact that an attack was underway. In a more recent event, attackers struck an unnamed steel mill in Germany by manipulating and disrupting control systems to such a degree that a blast furnace could not be properly shut down, resulting in ``massive''---though unspecified---damage~\cite{SicherheitDeutschland}.

\subsubsection{Anomaly Detection}

Given the relatively few examples of attacks targeting industrial control systems (compared to the vast number of attacks present in corporate networks), and the fact that adversaries may be well-funded nation-states that will not reuse attacks, the focus on intrusion detection for control systems has been on anomaly detection (%instead of
rather than signature-based detection).
Surveys of techniques related to learning and detection of anomalies in critical control systems can be found in~\cite{alcaraz_cazorla_fernandez,urbina2016survey}. 

Different kinds of anomaly-based IDS models have been suggested for SCADA systems~\cite{yang2006anomaly,atassi,chen,ye,hadziosmanovic2011,fovino2010modbus,Erez:2015:CVC:2822917.2823033}. Model-based anomaly detection for the Modbus/TCP industrial network protocol was first studied by Cheung et al.~\cite{cheung2007using}. They designed a multi-algorithm intrusion detection appliance for Modbus/TCP with pattern anomaly recognition, Bayesian analysis of TCP headers and stateful protocol monitoring, complemented with customized Snort rules~\cite{roesch}. Subsequent work~\cite{valdes2009communication} incorporated adaptive statistical learning methods for traffic patterns among hosts and traffic patterns in individual flows. Later, Briesemeister et al.~\cite{briesemeister2010detection} integrated these intrusion detection technologies into the EMERALD event correlation framework~\cite{emerald}.

Several of these learning-based models have difficulties in explaining the reasoning behind each alert. Sommer and Paxson \cite{sommer} discuss the surprising imbalance between the extensive amount of research on machine learning-based anomaly detection, and
%versus 
the lack of operational deployments of such systems. They argue that one of the reasons for this, is that 
%the 
machine learning-based anomaly detection systems 
%are lacking 
lack the ability to bypass the ``semantic gap'': The system ``understands'' that an abnormal activity has occurred, but it cannot 
%produce a message that will elaborate, helping 
help the operator differentiate between an abnormal activity and an attack. One of the goals of our anomaly detection tools is to produce models that not only detect suspicious activity, but that can inform operators why the activity is being reported and why it is important. 

\subsubsection{Automata-based models}
The periodicity of industrial control networks has been documented in several studies~\cite{barbosa,BarbosaPeriod,Kleinmann2017}. This periodicity can be captured by automata-based models, where requests and responses of industrial control networks are modeled as jumps between recurrent states. In addition, Automata-based models can be used to explain the behavior of  industrial networks to  operators. 

In one of the first papers on the topic, Goldenberg \& Wool \cite{Goldenberg201363} developed a model-based approach (the GW model) using a Deterministic Finite Automata (DFA) to represent the cyclic nature of the commands exchanged in Modbus traffic.  
Subsequently Kleinmann et al.~\cite{KleinmannWjdfsl} demonstrated that
this methodology is also successful in other network industrial protocols like Siemens S7. 
%a similar methodology is successful  in SCADA systems with Siemens S7 network traffic.

Caselli et al.~\cite{Caselli} proposed a probabilistic Discrete-Time Markov Chain (DTMC) model to capture sequences of SCADA messages. 
Based on data from three different Dutch utilities, the authors found that only 35\%-75\% of the possible transitions in the DTMC were observed. This strengthens the observations 
of a substantial sequentiality in SCADA communications~\cite{barbosa,Goldenberg201363,KleinmannWjdfsl}. However, unlike
\cite{Goldenberg201363,KleinmannWjdfsl} they did not observe clear cyclic message patterns. The authors hypothesized that the difficulties in finding clear sequences is due to the presence of several threads in the HMI's operating system that multiplex requests on the same TCP stream.

Kleinmann et al.~\cite{Kleinmann2017} 
introduced a modeling approach for such SCADA streams, using {\em Statechart DFAs}: the {\em Statechart} includes multiple DFAs, one per cyclic pattern. Each DFA is built using the learning stage of the GW model. Following this model, incoming traffic is de-multiplexed into sub-channels and sent to the respective DFAs. Kleinmann et al.~showed that if the correct DFAs are known, the {\em Statechart} model drastically reduces both the false-alarm rate and the learned model size in comparison with the naive single-DFA model. 
%%%%Not sure this citation fits the flow that we want to create in this section, therefore I commented it:
%However,  Kleinmann et al.~\cite{kawtbl2017} also  demonstrated that anomaly detection systems are blind to some types of attacks: specifically, the attacker may cause damage without changing the query patterns or timing by modifying the contents of data returning from PLCs, thus tricking the human operator into taking inappropriate and harmful actions. 

Recently, Faisal et al.~\cite{Faisal&al} analyzed a large data corpus from an industrial network, collected from a water facility in the U.S, and showed that the GW model performs quite poorly on this data set: only 28\% of the channels exhibited clear cyclic patterns. 
Following this research, Markman et al.~\cite{Markman2017} observed that data packets tend to be sent in bursts of packets, and that the bursts have an internal structure. Following this observation, Markman et al. suggested a new Burst-DFA model that learns the structure of the bursts, builds a directed graph for each channel (according to the order of the packets in the bursts), and provides a highly  accurate model of the traffic.

\subsection{Contributions}
Our starting point is the work of Markman et al.~\cite{Markman2017}. Our first contribution is a new method for ``fingerprinting'' traffic patterns 
%in a channel 
in a 
%certain 
given period of time. This enables us to introduce a measure of similarity between 
%the 
traffic patterns at different points in time. 

Our second contribution is demonstrating that the traffic patterns in the data channels have phases---there are stable traffic patterns followed by sudden changes 
%, and so forth. 
into a different but stable traffic patterns. 
This observation explains the deterioration of the average accuracy of previous models over time~\cite{Markman2017}. 

Furthermore, we introduce a method to automatically detect these traffic phase shifts. Our algorithm is based on clustering  the different time periods using our measure of similarity. This enables us to describe the traffic in a channel as a sequence of phases in time---potentially bridging (some of) the semantic gap that exists in anomaly detection models. 

Our fourth contribution is the introduction of an anomaly detection model that is phase aware, and the introduction of a sampling method to assure that the training set contains samples from the different phases in the channel. The sampling mechanism ensures that the model learns  traffic patterns from the different phases.
%in time. 

Finally, we introduce a permissiveness measure for the new model, developed by adjusting the measure of \cite{Markman2017}. %During the development of the level of permissiveness, we introduced a new method to count the number of unique paths in a union of k directed graphs. The method rely on set theory to count the elements of a union, along with the fact that counting the number of paths in an intersection of several directed graphs can be done using a proxy graph--- built by performing logical AND on the adjacency matrices describing the graphs, and applying paths counting operations on the proxy matrix.

We conducted an extensive evaluation of our new model on the water facility data corpus \cite{Faisal&al}, and our model, together with the sampling mechanism we suggested, improves the accuracy of previous models, while lowering the measure of permissiveness: i.e., the model has fewer false alarms, and is more specific.  

\section{Preliminaries}
\subsection{Adversary model}
We assume the adversary can take over the SCADA server or HMI, and issue control messages to devices in the field. The objective of the adversary is to manipulate the SCADA network to achieve an impact on the physical world.

% \begin{figure}[t]
% \centering
%   % \includegraphics[ width=0.65\textwidth,natwidth=500,natheight=60,
%   %trim = 100 340 20 -440]{Graphs/systemDiagram.eps}
%   \includegraphics[ scale = 0.08, trim = 40 20 20 20]{Graphs/systemDiagram-eps-converted-to}
%    \hfill
% \caption{Placing the Network Anomaly Detection System in a SCADA network}
%     \label{fig:systemDiagram}
% \end{figure}

Currently, most SCADA protocols do not include cryptographic algorithms such as ciphers and hash functions. Our adversary model assumes that if and when such security measures %shall be
are deployed, their associated cryptographic keys will be known to (or can be broken by) the adversary. 
We further assume that the adversary has in-depth knowledge of the architecture of the SCADA network and the various PLCs, as well as sufficient knowledge of the physical process and the means to manipulate it via the SCADA system. Thus the adversary has the ability to fabricate messages that would result in real-world physical damage.

One example of a semantic adversary is described by Fovino et al. for a system with a pipe in which high pressure steam flows \cite{fovino2010modbus}. The pressure is regulated by two valves. An attacker capable of sending packets to the PLCs can force one valve to close, and force the other to open. Each of these ICS commands is perfectly legal when considered individually, however when sent in an abnormal order they can bring the system to a critical state. In another example a system-wide water hammer effect is caused simply by opening or closing major control valves too rapidly~\cite{RobertT}. This can result in a large number of simultaneous damages.

Fundamentally these attacks work by injecting messages into the communication stream---possibly legitimate messages---on an attacker-selected pattern and schedule.
Hence a good anomaly detection system needs to model not only the messages in isolation but also their sequence and timing.

In our model, the network sensor needs to be located in a segment where it can passively monitor traffic that can be  modified by the adversary, and just before the PLC. The sensor is not located inline, so it does not affect the normal network operation (e.g., port mirroring or a similar mechanism is used to instruct the switch to send copies of the network traffic to the anomaly detection system).

Note that our anomaly detection approach does not distinguish between malicious events and faulty events. 

\subsection{The GW model}
The GW model \cite{Goldenberg201363} was developed and tested on Modbus traffic. Modbus is a simple request-response protocol widely used in SCADA networks.
A typical Modbus packet carries information about the \emph{message type}, the \emph{function code} specifying the service (e.g., read or write), and the \emph{memory address range} of data items. After the PLC processes the request, it sends a response back to the HMI.

In the GW model, the key assumption is that traffic is {\em periodic}, therefore, each HMI-PLC channel is modeled by a Mealy Deterministic Finite Automaton (DFA). The DFA for Modbus has the following characteristics:
(a) A symbol is defined as a concatenation of the \emph{message type, function code}, and \emph{address range}, totaling 33-bits;
(b) A state is defined for each message in the periodic traffic pattern.
The DFA represents the precise order of the symbols in the cyclic pattern. The GW model 
%suggests a network anomaly detection system that comprises 
has two stages: An unsupervised learning stage, and an enforcement stage. In the learning stage a fixed number of messages is captured (the learning phase assumes that the sniffed traffic is benign), the pattern length is 
%revealed,
identified, and a Mealy DFA is built for each HMI-PLC channel.
The channel's input-symbols are 
%categorized into 
divided in two groups: Known and Unknown. The Known group consists of all the input symbols that were observed during the learning phase, and have a matching DFA state. The Unknown symbols are those not observed in the learning phase. 
%all the rest.
In the enforcement stage, we monitor traffic in each channel, and trigger anomalies when the traffic is not recognized by its DFA. The model includes 3 types of anomalies: ``unknown'' symbol,  not seen during the training stage, ``miss'' for symbols that appear out of order, and ``retransmit'' symbols.

\subsection{The Burst-DFA model}
The Burst-DFA model \cite{Markman2017} was developed by Markman et al. The research was done by analyzing a corpus of Modbus traffic recorded at a large-scale water treatment plant in the U.S, which was previously found to be poorly modeled by cyclic-DFA models \cite{Faisal&al}. 
%%---->This seems a bit obvious in hindsight. This is the definition of a traffic flow; I guess a channel is just the continuous traffic flows 
The research found evidence of parallel TCP connections between the HMI and PLCs, 
%%led to refinements of the definition of a ``channel'' from the GW model, to 
so a refined definition of a ``channel'' includes TCP port numbers; in other words a channel is uniquely identified by the tuple (HMI IP, PLC IP, Unit-ID, PLC's port). 
A major finding of \cite{Markman2017} is that for each channel, the traffic exhibits bursty behavior---the HMI sends queries in bursts with defined construction, and with a relatively long time difference between consecutive bursts. Further, the research showed that the bursts have semantic meaning---the order within a burst depends on the messages.

Based on these observations a new model was suggested, which for each channel comprises a DFA that matches all the bursts of that channel, including their beginning and ending. For each channel, the burst-DFA learned the normal bursts one expect to see---the methodology allows automatic unsupervised learning of the individual channel traffic pattern. The training stage was done by dividing the data into channels, splitting each channel packet stream into bursts of data using the time difference between the packets, and building a directed graph in the form of an adjacency matrix.
In the enforcement stage, the model uses the burst-DFA to evaluate each data burst as it arrives, and ranks each query packet according to its position in the burst using the adjacency matrix. The Burst-DFA model uses the 3 anomaly indicators from the GW model (unknown, miss, retransmit), and adds two extra messages ``bad-beginning'' and ``bad-ending'' to indicate that a burst doesn't start or doesn't end with a symbol that was previously seen in these positions. 

The study introduced a metric to evaluate the permissiveness of the model---how strict or how general the model is. We discuss this metric in section 7, when we introduce our new model.

The burst-DFA model was evaluated on the water treatment data corpus. 
The Burst-DFA model successfully explains between 95\% to 99\% of the packets in the data-corpus, when the training set includes 50\% of the data.

\section{The Water Treatment Plant Data Corpus and Formal Definitions}
\subsection{Overview}
We used a one-day recording from a real-world operational large-scale water treatment plant in the U.S. Since Modbus is a Master/Slave protocol, we concentrate on modeling only the query packets by the HMI (and not add a state for the expected response). By only modeling the queries we can simplify the model, while not lowering the degree of generality. Modeling only the queries in a Master/Slave protocol is possible because for each query, there is a single response packet whose meta-data is fully determined by the query. Therefore modeling the responses does not add information in a DFA-based model that focuses only on the meta-data. 
Table~\ref{modelSizes} shows the statistics of the dataset. The packet-loss rate was calculated using a built-in wireshark tcp-analysis capability (\texttt{tcp.analysis.lost\_segment}).

\begin{table}[t]
\centering
\begin{tabular}{|l|c|c|}
\hline
\textbf{1}  & Total Packets    & 68,886,147                \\ \hline
\textbf{2}  & Duration              & 24h,3m,19s                    \\ \hline
\textbf{3}  & Packets Per Second             & 795                      \\ \hline
\textbf{4}  & Packet Loss            & 1.8\%                  \\ \hline
\textbf{5}  & \#IPs             & 99           \\ \hline
\textbf{6}  & \#Channels             & 935           \\ \hline
\end{tabular}
\caption{Data Corpus Statistics}
\label{modelSizes}
\end{table}

\subsubsection{Channel Separation and Identification}
In this research we follow the definition of a channel from \cite{Markman2017}, and so we use the 4-tuple (Master IP, Slave IP, Unit Identifier, Slave Port) to define a channel. By ``Slave Port'' we mean the source port that the HMI's TCP connection uses when sending a query---note that the ``Master Port'' is always 502 in Modbus.
We found 935 channels that exchange more than 500 packets  (covering 99.29\% of the packet capture). 
In the rest of the study, the numbering of the channels is arbitrary, made using MATLAB's ``Unique'' function over the 4-tupple (Master IP, Slave IP, Unit Identifier, Slave Port), before removing channels with less than 500 packets overall - this is why there are channel numbers above 935.

\subsubsection{Deterministic Finite Automata}
A classical DFA is a 5-tuple, $\left ( Q,\sum , \delta , q_{0}, F \right )$, consisting of: 
a finite set of states $\left (Q \right )$, a finite set of input symbols called the alphabet $\left (\sum \right )$, a transition function $\left (  \delta : Q \times  \sum \rightarrow  Q \right )$
, a ``start'' state $\left (q_{0} \in  Q \right )$ and a set of ``accept'' states $\left (F \subseteq Q \right )$.

\subsubsection{Input Symbols and States}
We follow the definitions of the symbols from the GW model and the Burst-DFA model:\\
The states that are reached after a query message are called Q-states. Respectively, states that are reached after response messages are called R-states.
We have chosen to only model the sequence of queries in each channel due to the fact that the communication is in a Master-Slave protocol.
A Modbus query consists of the following fields: Transaction Identifier $\left (T.ID\right )$, Function Code $\left (FC\right )$, Reference Number $\left (RN\right )$, and bit/word count $\left (Count\right )$.
We define a symbol in the alphabet as a 3-tuple $(FC, RN, Count)$. 
We say that a symbol is a known-symbol if it appears in the training set of the particular channel, and an unknown-symbol otherwise.
For each symbol $s_{i}$, we define a state $S_{i}$, as the DFA state following the occurrence of the symbol.

\section{Burst-DFA model performance over time}
\begin{figure}[t]
\centering
  \includegraphics[ scale = 0.4, trim = 40 20 20 20]{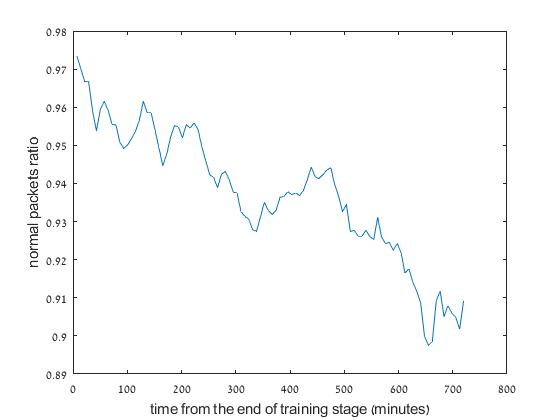}
   \hfill
\caption{Fraction of normal packets Vs time from the training stage}
    \label{fig:normalVsTime}
\end{figure}

\begin{figure}[t]
\centering
  \includegraphics[ scale = 0.4, trim = 40 20 20 20]{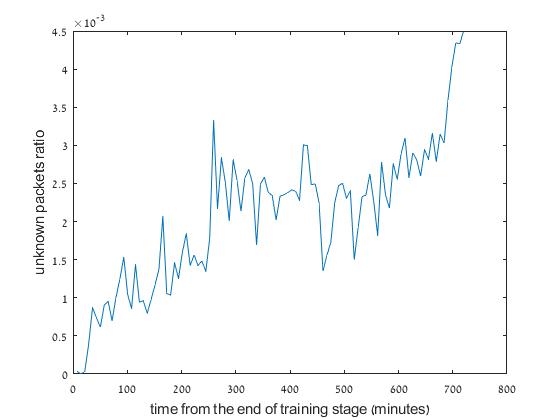}
   \hfill
\caption{Fraction of unknown packets Vs time from the training stage}
    \label{fig:unknownVsTime}
\end{figure}

In the Burst-DFA model \cite{Markman2017}, the training was done on the first 50\% of the data corpus, and the testing was done on the remaining 50\%. When evaluating the performance of the model, we noticed that the accuracy deteriorates over time---on average the model describes the traffic that came right after the training data better than later traffic.
Figures \ref{fig:normalVsTime} and \ref{fig:unknownVsTime} show the percentage of the ``Normal'' packets (packets that the model explains correctly) and ``Unknown'' packets (packets that the model sees for the first time during the test stage) every 7 minutes, as a function of time since the end of the training data.

Even though the overall accuracy of the model is good---when looking at the accuracy Vs time in figure \ref{fig:normalVsTime}, it is easy to see the deterioration.

\subsection{The source of the decline in accuracy}
A possible reason for the deterioration in accuracy could be a steady decline in accuracy over time in many of the channels. An alternate hypothesis is that the changes are abrupt. % changes in accuracy in different channels. 
In order to check  the validity of the abrupt-changes hypothesis, we searched for channels in which the model experienced sudden changes in false alarm rate.

\subsection{Performance vs time - evidence of traffic phases}
In order to find examples of channels where the traffic exhibits phases, we divided the data set into 100 equal parts of 15 minutes each, and measured the performance of the Burst-DFA model in each part separately. We then went through all of the 935 channels, and searched for channels in which in some part of the data set, one of the anomaly indicators of the Burst-DFA model goes above 20\% (unknown, miss, retransmit, bad beginning/ending of a burst). This somewhat naive search highlighted 180 suspicious channels out of the total 935. Visually inspecting some of the channels showed sudden changes in the model's accuracy over time, in a way that indicates a sudden change in the traffic pattern. 
\begin{figure}[t]
\centering
  \includegraphics[ scale = 0.4, trim = 40 20 20 20]{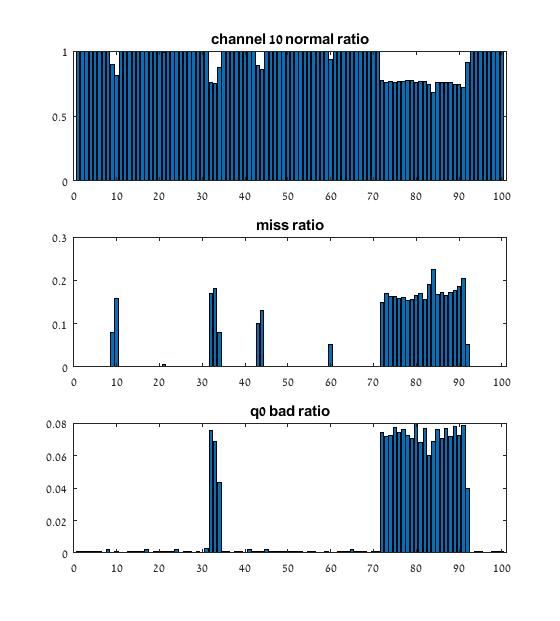}
   \hfill
\caption{From top to bottom: Channel 10 Fraction of Normal Packets, Misses and Wrong Beginning vs time}
    \label{fig:channel_10_indicators}
\end{figure}
For example, Figure \ref{fig:channel_10_indicators} shows the normal ratio (accuracy), misses ratio and the number of bursts starting with the ``wrong'' state in channel 10. The Figure shows that the change in accuracy is very sudden, and that the accuracy returns to the higher values after about 20 time units. We observed similar phenomena in many other channels---this clearly indicates that indeed there are ``phases'' in the traffic---at different times, the traffic  follows different rules and regularities.

\section{Traffic phases} After observing the existence of traffic phases by looking at the accuracy level of the Burst-DFA model over time, we turn our attention to detecting traffic phases directly, without using the result of the Burst-DFA model as a proxy. We define ``traffic phases'' as parts of the packet stream in a specific channel that exhibit a distinct pattern or a distinct set of possible packets. %We will search for such parts using the Adjacency matrices that were developed in \cite{Markman2017}, the matrices that describe the traffic pattern of a certain channel by including the different possible packets in the channel, and the possible transitions between those packets in the data bursts. We search for the different phases by looking at some similarity indexes between the adjacency matrices of the channel in different times - if two adjacency matrices are similar, they describe similar traffic patterns, and if they vary significantly, the traffic patterns are different and therefore the phases are different.

\subsection{Using Adjacency Matrices to identify phases in the data}
The Burst-FDA model \cite{Markman2017}  builds an adjacency matrix for each channel, incorporating the different possible bursts of data---the possible set of packets and the possible order of the packets. An adjacency matrix is a description of a Deterministic Finite Automata (DFA), in which the nodes are the rows and columns of the adjacency matrix (unique packets), and the values of the matrix are the number of transitions between the nodes (the frequency of packets). In the next sections we describe the process of building the adjacency matrices, and introduce a measure of similarity (between adjacency matrices), that allows us to divide the network traffic stream into different segments,  similar to each other. If two adjacency matrices are similar, they describe similar traffic patterns, and if they are very different,  the traffic patterns are dissimilar, and therefore the traffic belongs to different phases.
\subsubsection{Building the adjacency matrix}
We begin by separating the traffic into different channels as described in section 2. We then divide the packet stream in each channel to bursts of data, as described in \cite{Markman2017}. Algorithm \ref{alg:adj_mat} describes the construction of the DFA's vertex set and transition function (as an adjacency matrix).

\begin{algorithm}[t]
  \caption{Adjacency Matrix Formation
    \label{alg:adj_mat}}
  \begin{algorithmic}[1]
    \Require{$B$ - a set of $k$ Bursts with $n$ unique symbols}
    \Statex
%    \Comment{$\oplus$: bitwise exclusive-or}
    \Function{adjMat}{$B$}
    \Let{$adj\_mat$}{$\{0,...,0\}$}
	\Let{$n$}{\#(unique symbols in the channel)}
	\Let{$k$}{\#(number of bursts in $B$)}
	\Let{$uStates$}{\{unique symbols in channel,$q_{0},q_{end}$\}}
      \For{$i \gets 1 \textrm{ to } k$}
      \Let{$l$}{$length(burst(i))$}
       \Let{$curr\_state$}{$q_{0}$}
       \For{$j \gets 1 \textrm{ to } l$}
       \State $adj\_mat(curr\_state,S_{j})$++
        \Let{$curr\_state$}{$S_{j}$}
       \EndFor  
       \Let{$adj\_mat(curr\_state,q_{end})$}{1}
      \EndFor
      \State \Return{adj\_mat}
    \EndFunction
  \end{algorithmic}
\end{algorithm}

The resulting adjacency matrix represents a DFA, based on the bursts given to the algorithm.
We add an $\varepsilon$-transition to $q_{end}$ at the end of each burst to mark the burst ending.

\subsubsection{Computing the similarity between adjacency matrices}
Given two adjacency matrices, we wish to calculate a score measuring their similarity. To do so we first reshape each adjacency matrix into a vector shape. We then normalize each vector to unit length by dividing it by its magnitude (using the $L^2$ norm), to ensure the similarity measure is based on the transitions frequencies and not on the absolute number of transitions. We now have two normalized vectors, and we can use one of many methods to compute their similarity. We tested the Euclidean distance between vectors, and the correlation between vectors as possible measures of similarity, and the two options gave similar results. From now on we only discuss the Euclidean distance measure (a small distance means similar adjacency matrices).

\subsubsection{Using adjacency matrices to describe the data}
We use the idea of the adjacency matrices as descriptors of the data stream in order to find the different phases in the traffic. We use the following procedure:
\begin{enumerate}
  \item Divide the data into channels. 
  \item For each channel, separate the data stream into bursts of packets as in \cite{Markman2017}.
  \item Separate the list of bursts into 100 equal parts.
  \item Use algorithm \ref{alg:adj_mat} to build an adjacency matrix of size $sXs$ for each part, when s is the number of unique symbols in the channel.
  \item Vectorize and normalize each matrix.
  \item Compute the Euclidean distance between each pair of vectors.
\end{enumerate}

\subsection{Specific channel examples}
\begin{figure}[t]
\centering
  \includegraphics[ scale = 0.4, trim = 40 20 20 20]{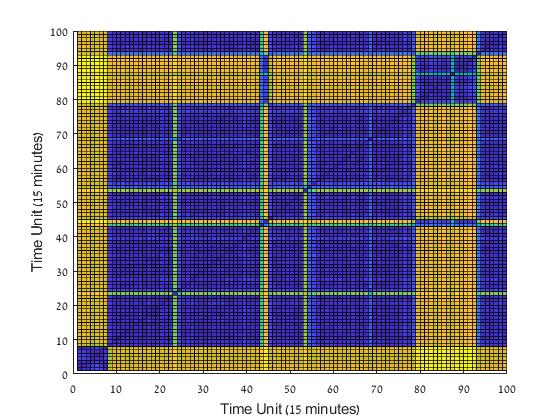}
   \hfill
\caption{Channel 10 distance between adjacency matrices from different times, the blue (dark) color indicates low distance, and the green high distance}
    \label{fig:channel_10_distance}
\end{figure}

\begin{figure}[t]
\centering
  \includegraphics[ scale = 0.4, trim = 40 20 20 20]{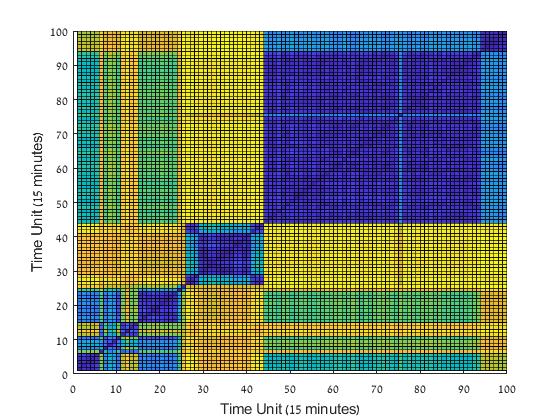}
   \hfill
\caption{Channel 130 distance between adjacency matrices from different times}
    \label{fig:channel_130_distance_matrix}
\end{figure}

After calculating the chosen similarity measure between each pair of the 100 adjacency matrices, we plotted the distance matrix for the channels and inspected them. For instance, Figures \ref{fig:channel_10_distance} and \ref{fig:channel_130_distance_matrix} show the distance matrices between the 100 parts of channels 10 and 130. 

As can be seen from Figure \ref{fig:channel_10_distance}, in channel 10 there are 3 main traffic phases: 
\begin{enumerate}
\item from the first time unit to the 8\textsuperscript{th} time unit.
\item from the 9\textsuperscript{th} time unit to the 80\textsuperscript{th} time unit except for the 44\textsuperscript{th} and 45\textsuperscript{th} time units. The same phase reappears from the 91\textsuperscript{st} time unit to the end of the recording.
\item The 44\textsuperscript{th} and 45\textsuperscript{th} time units, and from the 80\textsuperscript{th} time unit to the 91\textsuperscript{st} time unit.
\end{enumerate}
Figure \ref{fig:channel_130_distance_matrix} shows a qualitatively similar situation in channel 130: we can visually identify $6-9$ distinct phases.
From the two examples shown above, it is clear that there are traffic phases, and we can see that it is easy to visually detect the transition between the traffic phases. We can learn from the two examples that the phases can last for several hours, and we can see that some phases appear only at the end of the recording.

\subsection{Intermediate conclusions from the examples}
Once we realize that channels exhibit phases in their traffic, we can draw some conclusions: First, in order to model the traffic pattern correctly, we need our training data to include patterns from all of the traffic phases. It is clear from the example in Figure \ref{fig:channel_10_distance} that if the training set in channel 10 consisted of the first 30\% of the data, the training would have missed the 3\textsuperscript{rd} phase altogether, and the accuracy of the model would have deteriorated. Second, if the traffic patterns have phases, we can use this fact to create a better model - a model that still has high accuracy, and that is less permissive than the Burst-DFA model. 
\section{analyzing the traffic phases}
When looking at Figure \ref{fig:channel_10_distance}, we can visually detect the different phases. We would like to detect the phases automatically in order to check whether or not the traffic in a certain channel is divided into phases, and in order to build a traffic model. 

We want to segment the time series into different phases; in order to do that we organize the time parts into clusters based on the similarity between their adjacency matrices. To achieve this time series segmentation, we use $k$-means clustering.
%, such that similar adjacency matrices will be in the same cluster. In order to do so we used the $k$-means clustering algorithm.
\subsection{$k$-means clustering}
$k$-means is a classical clustering algorithm which aims to partition $n$ observations (vectors) into $k$ clusters, in which each observation belongs to the cluster with the nearest mean, serving as a prototype of the cluster \cite{lloyd1982least}. The mean of a cluster is defined as a vector in which the value of each coordinate is calculated by averaging the values of the same coordinate in all of the vectors in the cluster, i.e., calculating a coordinate-wise average of the vectors in the cluster. The input to the algorithm is a set of $n$ samples, and a predefined number of clusters---$k$. The algorithm works iteratively, when in each iteration every sample is assigned to the cluster whose $mean$ is closest to it, and the mean of each cluster is updated according to the updated members of the cluster. The algorithm keeps iterating until cluster assignments do not change, or the maximum number of iterations is reached (we used the default number in matlab---100 iterations). The initial centers of the clusters are selected randomly, and so running the algorithm multiple times with the same input may produce different results. In our case we use the clustering algorithm on the 100 adjacency matrices in each channel, and we treat the output clusters as the phases of the channel---each cluster the algorithm outputs is a set of all of the times in the recording with similar traffic pattern. When using $k$-means clustering, it is a challenge to choose the parameter $k$. We next bring an example of the result of the clustering algorithm in a specific channel, and then we describe our mechanism for choosing $k$ automatically.

\subsubsection{Example}
To demonstrate the results of the clustering algorithm, we continue with channel 10 we discussed previously. By looking at Figure \ref{fig:channel_10_distance} we observed that there are 3 main phases, so in this example we manually chose $k$=5.
\begin{figure}[t]
\centering
  \includegraphics[ scale = 0.4, trim = 40 20 20 20]{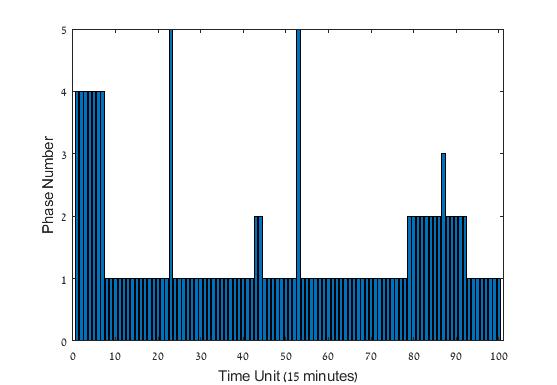}
   \hfill
\caption{$k$-means cluster analysis of channel 10: the $X$ axis is the time slot $t$, the $Y$ axis is the number of the cluster that is assigned to the adjacency matrix at time $t$.}
    \label{fig:channel_10_clustering}
\end{figure}
We can see in Figure \ref{fig:channel_10_clustering} that the algorithm assigned the phases in a good fit to our observation: there are 3 main phases, and the transitions between the phases appear at the times we anticipated. After convincing ourselves that the clustering mechanism describes the phases correctly in various channels, we turn to the task of automatically choosing the number of clusters---$k$.

\subsection{Choosing $k$ - the number of clusters}
Determining the number of clusters in a data set is a fundamental issue in clustering algorithms such as $k$-means, in which the user needs to set the value of $k$.
While there are many different methods of choosing $k$, there is no definitive answer to this problem \cite{charrad2012nbclust} \cite{pham2005selection}.
It is clear that the optimal value of $k$ depends on the distribution of the data, but it also depends on the similarity measure used, and even on the user resolution preferences - accuracy versus generality of the model. In the extremes, choosing $k=1$ results in a single cluster with high variance between the elements of the cluster, and choosing $k=n$ results in a cluster for each data point and no generality. The choice of $k$ needs to reach a balance between these extremes. We chose to use the silhouette method of determining $k$ \cite{rousseeuw1987silhouettes}. The silhouette of a data instance (in our case, an adjacency matrix) is a measure of how closely it is matched to other instances within its cluster and how loosely it is matched to data instances of the closest neighboring cluster---the cluster with the closest mean, i.e., the cluster whose mean is closest to the instance when using $L^2$ distance. By averaging over all of the data samples it is possible to appreciate the relative quality of the clusters. 
The silhouette value for the $i^{th}$ instance, $S(i)$, is defined as
\begin{equation}S(i) = \frac{(b(i)-a(i))}{max(a(i),b(i))}\end{equation}
where $a(i)$ is the average distance from the $i^{th}$ instance to the other instances in the same cluster as $i$, and $b(i)$ is the average distance from the $i^{th}$ instance to instances in the closest cluster.
The silhouette value ranges from $-1$ to $+1$, where a high silhouette value indicates that the data instance is well-matched to its own cluster, and poorly-matched to neighboring clusters. If most instances have a high silhouette value, then the clustering solution is appropriate and vice versa.
The silhouette is computed after performing the clustering, so choosing the ``best'' $k$ is done by trying different values in different clustering iterations, and choosing the value of $k$ that gives the best results---the highest average silhouette value.
\begin{figure}[t]
\centering
  \includegraphics[ scale = 0.4, trim = 40 20 20 20]{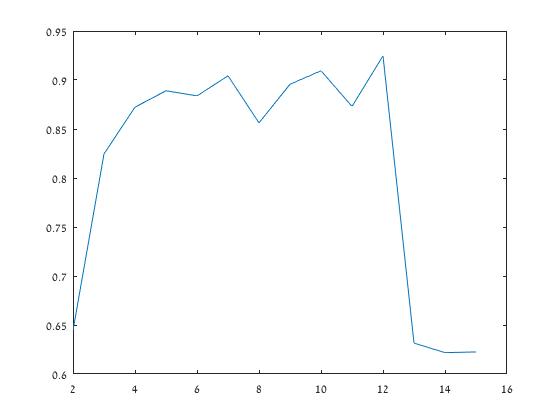}
   \hfill
\caption{Channel 130 silhouette graph for $k$ between 1 and 15. Each point represents the average silhouette value over all 100 adjacency matrices for a single run of $k$-means.}
    \label{fig:channel_130_silhouette}
\end{figure}

When applying the silhouette method on real data, we noticed some variation in the results that prevents us from just choosing the $k$ value that gives the maximum silhouette. Since the clustering algorithm includes a random seed, the results of running the $k$-means algorithm twice with the same $k$ and the same data are usually not equal, and so the silhouette values vary as well. Figure \ref{fig:channel_130_silhouette} shows the silhouette of channel 130, calculated on $k$ values between 1 and 15. We can see that the the graph is noisy---the silhouette value of $k=7$ and $k=9$ are greater then the value for $k=8$, which indicates noise since we wish to find the ``natural'' number of clusters in the data. This noise is the result of the random nature of the algorithm and the distribution of the data. Furthermore, we can learn from Figure \ref{fig:channel_130_silhouette} that in fact there is no clear optimal value for $k$---the maximum value in the graph is for $k=12$, but the values for $k=5$ up to $k=11$ are similar, and therefore may be appropriate. Understanding that the $k$-choosing task is ambiguous, even when using a known scientific method, we chose to to search for the maximum value of $k$ 3 times, and chose the minimum ``optimal'' $k$ we got. 

\subsection{Phase shift analysis}
Once we have a clustering method to identify the phases, we can study the behavior of the phases in the different channels. We found that some of the channels exhibit clear multi-phase behavior as seen in channels 10 and 130 described above (recall Figures \ref{fig:channel_10_distance} \ref{fig:channel_130_distance_matrix}), but other channels do not exhibit any such behavior---the distance matrix doesn't look like a block matrix, and there are no continuous phase patterns. 
\begin{figure}[t]
\centering
  \includegraphics[ scale = 0.4, trim = 40 20 20 20]{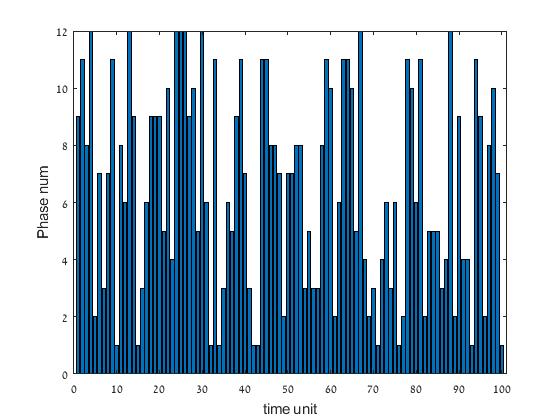}
   \hfill
\caption{Channel 60 phases in time - there is no continuous phase patterns, with $k=12$ possible phases}
    \label{fig:channel_60_clustering}
\end{figure}

Channel 60 is an example of such a channel (see Figure \ref{fig:channel_60_clustering}). The optimal $k$ value was 12, but the assignment of phases to adjacency matrices shows no clear temporal continuity; there are many phase shifts, which contradict our intuition of what a ``phase'' means. 

\begin{figure}[t]
\centering
  \includegraphics[ scale = 0.4, trim = 40 20 20 20]{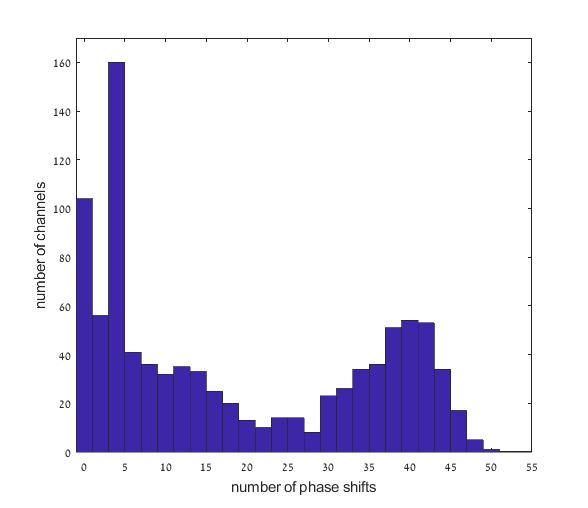}
   \hfill
\caption{Histogram of the number of phase shifts across all channels}
    \label{fig:phase_shifts_in_channels_histogram.jpg}
\end{figure}

After running the clustering algorithm on all channels, we counted the number of phase shifts along the recording in order to describe the overall behavior of the channels. Intuitively, channels with only a few phase shifts exhibit phases, and channels with many phase shifts (like channel 60) exhibit no such behavior. Figure \ref{fig:phase_shifts_in_channels_histogram.jpg} shows the histogram of the number of phase shifts over all of the channels. From the distribution we can learn that  44\% of the channels have 10 or less phase shifts---those are channels that exhibit some kind of multi-phase behavior. In fact the highest peak is at 5, showing that around 140 channels exhibited 5 phase shifts over 24 hours: a very reasonable number. Conversely, 38\% of the channels have more than 25 phase shifts, i.e., our analysis showed no multi-phase behavior. Possibly some of the channels we found to have many phase shifts may still have phases, but our model didn't fit these channels well enough (maybe other choice of hyper-parameters or model would disclose multi-phase behavior). 

We argue that automatic phase-shift detection may be of value beyond anomaly detection in the traffic. We believe that a traffic phase lasting many hours probably has semantic meaning that is correlated with a phase in the controlled process. As such, it may be labeled by the process engineer (e.g., as ``Chlorinating''/``Mixing'' etc.) during the training period, and displayed visually during the model enforcement. This type of semantic labeling is similar to the approach of \cite{Fauri:2017:SSA:3140241.3140250} and may assist in reducing the semantic gap \cite{sommer2010outside}. 

To illustrate the connection between the phase shifts and the controlled process, we analyzed a particular phase shift in channel 130 at the Modbus packets level. As can be seen in Figure \ref{fig:channel_130_distance_matrix}, there is a phase shift after about 25\% of the recording---about 6 hours into the recording. When looking at the bursts assignment into phases (as described in section 7), a new set of traffic patterns appears at that time. The set of patterns includes a new query that was previously unseen---a request to read the 2 bytes from a new register range (Modbus Reference Number)---1186. This example implies that the phase shifts are actually connected to a change in the operation of the controlled system, and hence has the potential to bridge some of the semantic gap.

\section{A Phase aware anomaly detection model}
After witnessing the existence of traffic phases, we suggest a model that incorporates the phase detection capability, to provide high accuracy while limiting the permissiveness of the model. 
Our model creates a few sub-models during the training stage, according to the different traffic phases, and checks to see if the data fits one of the sub-models during the enforcement stage. This is a reminiscent of the state-chart approach of \cite{KleinmannWjdfsl}, but with a different construction method.
We begin by discussing the data collection process for the training set, as the usual process of separating the data set into training and test set by time may prove insufficient.
\subsection{Training set assembly via burst-based sampling}
As seen in Figure \ref{fig:channel_10_clustering}, traffic phases may only appear after a long time from the beginning of the recording---more than 10 hours in this example. It is important to include samples from all traffic phases in the training set in order to learn all possible patterns and to have an high accuracy model. In the Burst-DFA model \cite{Markman2017} this resulted in taking the training set to be 50\% of the data, which resulted in an accurate but highly permissive model. We suggest another option to assemble the training set---burst-based sampling. 
The main idea is to collect every $n$\textsuperscript{th} burst of data and include it into the training set, while skipping the rest. This way we can have a long training period while reducing the amount of traffic we need to gather, and still include all traffic phases. It is important to emphasize that we sample bursts of data, and not individual packets. If we would sample the packets instead we would lose the ability to use the internal structure of the bursts, and the order of the packets in the pattern. We use the sampling mechanism to demonstrate the importance of the inclusion of all phases in the training set. To evaluate the impact of burst-based sampling on a fixed data corpus, we train the model on a non-**** sample of $\frac{1}{n}$ of the bursts, and test the model on the remaining $(1-\frac{1}{n})$ of the bursts.

\subsection{The training stage}
In the training stage we begin forming a training set of bursts according to the steps discussed in section 5 and the burst-based sampling method. By repeating the steps from sections 5 and 6 on the training set we form 100 adjacency matrices, each labeled by a number from 1 to $k$, according to the cluster it is in. We now combine all of the adjacency matrices in each cluster using logical OR (we ignore the frequency of transitions), and we remain with $k$ adjacency matrices (each a union of adjacency matrices from a cluster) representing the model---k DFA's in total. The model is different from the Burst-DFA model \cite{Markman2017}, that was made of one big adjacency matrix, formed as a union of the 100 adjacency matrices from our model. 

\subsection{The enforcement stage}
Given $k$ adjacency matrices, representing the $k$ DFA's for a channel, we can compare the channel's traffic to the model, and flag anomalies in the enforcement stage. 
Similarly to the model presented in \cite{Markman2017}, we evaluate finite bursts---we move through the adjacency matrices from the starting state $q_{0}$ of each burst, and ensure we reach $q_{end}$ at the end of the burst. Unlike \cite{Markman2017}, each burst of data is compared to all $k$ DFA's from the training stage, in order to check for anomalies, and in order to determine to which traffic phase the burst belongs.

\subsubsection{The transition tunction for a single adjacency matrix}
The transition function in a single DFA is a transformation that for each $\left(Base State, Input Symbol\right)$ tuple
returns a ($Dest State$, $Operation$) tuple. The transition function implements the behavior predicted by the model. According to previous studies we define six types of transitions, one Normal, and five anomalous: Unknown, Miss, Retransmit, Wrong-Beginning and Wrong-Ending (see \cite{Goldenberg201363} \cite{Markman2017} for details). 

\subsubsection{Single burst phase assignment}
Each burst of data in the test stage is compared to all $k$ DFAs from the training stage---for each one a vector of the counters of the different categories is formed \textit{(Normal, Miss, Unknown, Retransmit, Wrong-Beginning, Wrong-Ending)}. We choose to assign the burst to one of the $k$ phases by choosing the adjacency matrix that minimizes the Unknowns for the burst. If more than one adjacency matrix gives the same minimal number of unknowns, we take the one that results in the lowest number of overall anomalies \textit{(Miss, Retransmit, Wrong-Beginning, Wrong-Ending)}. The output for each burst is the phase it is assigned to, and the vector of transition function categories.

\subsubsection{Enforcement stage output}
The enforcement stage of a particular channel is made by evaluating all of the bursts in the test set according to the $k$ DFA's provided by the training stage.
For each channel, the result of the enforcement stage is a vector of counters of the different transition function categories:
Normal, Miss, Unknown, Retransmit, Wrong-Beginning, Wrong-Ending. The values are the sum of the counters for all of the bursts in the channel's test set.
Algorithm \ref{alg:Analyze Single Burst} summarizes the implementation of the enforcement process on each burst.

\begin{algorithm}[t]
  \caption{Analyze Single Burst
    \label{alg:Analyze Single Burst}}
  \begin{algorithmic}[1]
    \Require{Denote the symbols of the burst by \{$s_1$,...$s_b$\}}
    \Statex
%    \Comment{$\oplus$: bitwise exclusive-or}
    \Function{analyzeSingleDFA}{$adj\_mat, burst$}
    \Let{$Normal\_Counter$}{$0$}
	\Let{$Miss\_Counter$}{$0$}
	\Let{$Unknown\_Counter$}{$0$}
	\Let{$Retransmit\_Counter$}{$0$}
	\Let{$Wrong\_Beginning\_Counter$}{$0$}
	\Let{$Wrong\_Ending\_Counter$}{$0$}
	\If {(adj\_mat($q_{0} , S_{1}$)$>$0} %AW : OK
	\State Normal\_Counter++
	\Else
	\State Wrong\_Beginning\_Counter++
	\EndIf 
      \For{$i \gets 1 \textrm{ to } (burst\_len-1)$}
      \Let{$current\_state$}{$S_{i}$}
      \Let{$next\_symbol$}{$s_{i+1}$}
        %%\If{$current\_state \mbox{ is unknown}$}
        \If {($next\_symbol \mbox{ is unknown}$)}
          \State Unknown\_Counter++
	 \ElsIf {($current\_state$ = $S_{i+1})$}
	   \State Retransmit\_Counter++
 	\ElsIf{$(adj\_mat(S_{i}, S_{i+1})$}
	   \State Normal\_Counter++
 	\Else
	   \State Miss\_Counter++
        \EndIf
      \EndFor
	\If{$(adj\_mat( S_{b-1}, q_{end})>0)$}
	\State  Normal\_Counter++
	\Else 
	\State Wrong\_Ending\_Counter++
	\EndIf
      \State \Return{All Counters}
    \EndFunction
    
    \Function{analyzeBurst}{$adj\_mat[k], burst$}
    \Let{$res\_vec[k]$}{$0$}
    \Let{$res\_for\_order[k][2]$}{$0$}
    \For{$i \gets 1 \textrm{ to } k$}
    	\Let {$res\_vec[i]$}{analyzeSingleDFA($adj\_mat[i], burst$)}
        \Let{$res\_for\_order[i][1]$}{$unknown$}
        \Let{$res\_for\_order[i][2]$}{$sum(anomaly\_indicators)$}
    \EndFor
    \Let {index}{sort($res\_for\_order$,1,2)}
     \Comment{sort by the first column and then by second column}
    \State \Return{index, $res\_vec[index]$}
    \EndFunction
  \end{algorithmic}
\end{algorithm}

\subsection{Analyzing the Permissiveness of the Model}
It is important to understand how constrained or how permissive our model is. In the extreme, a channel with a single burst pattern will generate a single linear DFA with only one path from $q_{0}$ to $q_{end}$---a very constrained model. Conversely, if all $s$ symbols are observed in every one of the $b$ positions in the burst, then the model will allow all $s^{b}$ paths through it---a permissive model. In \cite{Markman2017} we introduced the $R_{perm}$ measure to describe the level of permissiveness of the model. The permissiveness measure is a normalized ratio between the number of paths the model ``allows'', and the number of potential paths allowed through the most permissive model. We need to further develop this measure to incorporate the structure of the new model. 
\subsubsection{Permissiveness of a single DFA - $k$=1}
This is a recap from \cite{Markman2017}. Conveniently, calculating the number of paths of length $l$ in a directed graph, using one adjacency matrix, is well known: if $A$ is the adjacency matrix, then after raising $A$ to the power of $l$, $A_{ij}^{l}$ counts the number of paths of length $l$ from vertex $i$ to vertex $j$.
In our model, each burst starts with $q_{0}$ and ends with $q_{end}$, and in between there are $b$ states, where $b=burst\_length$. Therefore, the number of paths from 
 $q_{0}$ to $q_{end}$ when $k$=1 can be calculated by looking at the $ (q_{0},q_{end})$ cell in the matrix $A^{(b+1)}$. We raise to the power of $b+1$ to allow for the $\varepsilon $-transition edges to $q_{end}$.

With this, we define the permissiveness of a single DFA, for burst-length $b$ over $s$ symbols to be:\begin{equation}R_{perm}=\sqrt[b]{\frac{ \#\mbox{allowed-paths} }{ s^{b} }}=\frac{\sqrt[b]{ \#\mbox{allowed-paths}}}{ s}
\end{equation}
In \cite{Markman2017} we showed that: \begin{equation}\frac{1}{s} \leq R_{perm} \leq 1\end{equation}
When $R_{perm}$ of the model in a specific channel is close to the lower bound, we can say that our model is very constrained, and when $R_{perm}$ is close to 1 we would say that the model is permissive.
 
\subsubsection{Extension to $k$ DFAs - calculating the number of unique paths in a group of graphs}
In order to expand the definition of $R_{perm}$ to our case, we need to change the way we count the number of paths allowed by our model. We can count the number of paths in any single DFA (adjacency matrix) using the method described above, but simply adding the number of the paths from the $k$ DFAs is insufficient since some paths (bursts) may be allowed in more than one DFA, which will result in over counting. A naive approach is to simply list all of the possible paths from each sub-model, and then count the unique paths, but this can be very inefficient, up to $ \Omega (k\cdot s^b)$ work. We would like to use graph theory and set theory in order to complete this task. Suppose $k=2$, and we have 2 DFAs, we can count the number of paths in each sub model and sum the results, but we need to subtract the paths that appear in both DFAs. We perform this task using the fact that the paths that exist in both graphs appear in the intersection of the two graphs. We can get the graph that represent the intersection between the two graphs by performing logical AND on the two adjacency matrices describing the graphs, and the resulting matrix represents a graph with all of the nodes, and only the edges that appear in both sub-model. We can generalize this method into every value of $k$ using De Morgans laws of inclusion-exclusion principle: \begin{equation}\left | \bigcup_{i=1}^{k}A_i{} \right | = \sum_{\phi \neq J\subseteq \left \{ 1,2,...,k \right \}}(-1)^{\left | J \right |-1}\left | \bigcap_{j\in J}A_j \right | \end{equation} where the intersection calculation is performed using the generalization of the adjacency matrices intersection for $k>2$. This approach works faster than the naive method when $k$ is relatively small and $s^b$ is large. In our experiments we restricted ourselves to $k\leq 15$. 
\section{Model Results}
We introduced two main concepts in our model: 1. Burst-based sampling the training data (section 7). 2. Using $k$ DFAs to evaluate the bursts in each channel. In this section we present the results in comparison with the Burst-DFA model \cite{Markman2017}, and we show the results with and without the sampling mechanism. The k-phase model permits fewer transitions between states compared to the Bursrt-DFA model, So given the same training set, we expect the accuracy of the model to be slightly worse than that of the Burst-DFA model. However, we expect the new model to be less permissive. Having said that, the sampling mechanism should improve the accuracy of the model, so a success will be a combination of the sampling mechanism and the new model thats result in better accuracy and less permissiveness than the Burst-DFA model without sampling. We demonstrate the results using a training set that is 33\% of the data set. 
We describe the accuracy of the model by the percentage of normal queries: \begin{equation} \frac{normal}{total~queries} \end{equation} And the percentage of queries that are either normal, miss, or retransmit:\begin{equation}  \frac{normal+miss+retransmit}{total~queries} \end{equation}
In addition to these measures of success, we also checked the Wrong-Beginning and Wrong-Ending ratio (out of all bursts), to understand how well we model the structure of the bursts.

\subsection{Model Accuracy}
Table \ref{table:results_summary} summarizes the accuracy results of the k-phase model compared to the previous Burst-DFA model. We can learn from the table that using burst-based sampling mechanism drastically improves the accuracy of both models, and that the k-phase model is slightly less accurate than the Burst-DFA model when using sampling. If we do not use sampling, the Burst-DFA model is more accurate than the k-phase model.
\begin{table}[]
\centering
\caption{Results Summary}
\label{table:results_summary}
\begin{tabular}{|l|l|l|}
\hline
\textbf{model\textbackslash{}accuracy ratio}                                  & \textbf{normal  ratio} & \textbf{\begin{tabular}[c]{@{}l@{}}normal+miss+\\ retransmit ratio\end{tabular}} \\ \hline
\textbf{k-phase}                                                            & 88.2\%                 & 99.7\%                                                                           \\ \hline
\textbf{k-phase+sampling}                                                   & 98.9\%                 & 99.99\%                                                                          \\ \hline
\textbf{Burst-DFA}                                                      & 92.8\%                 & 99.7\%                                                                           \\ \hline
\textbf{\begin{tabular}[c]{@{}l@{}}Burst-DFA +sampling\end{tabular}} & 99.0\%                 & 99.99\%                                                                          \\ \hline
\end{tabular}
\end{table}

\begin{figure}[t]
\centering
  \includegraphics[ scale = 0.4, trim = 40 20 20 20]{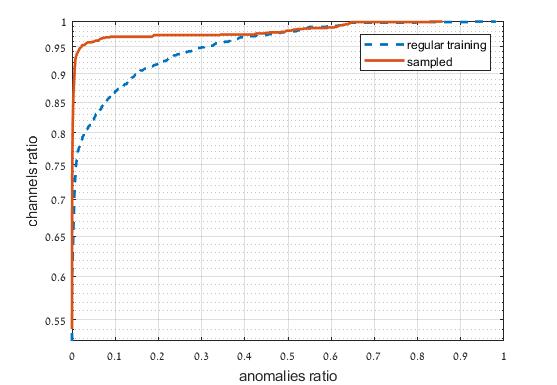}
   \hfill
\caption{The total anomalies CDF for our model over all of the channels, with and without training set sampling}
    \label{fig:anomalies_CDF}
\end{figure}

Table \ref{table:burst_structure_results} summarizes the percentage of \textit{bad beginning} and \textit{bad ending} of the k-phase model, compared to the Burst-DFA model. We see that using burst-based sampling improves the accuracy in those two parameters dramatically, and that the difference between the models when using sampling is less significant than the difference when not using sampling. When inspecting the difference between the accuracy of the model with or without sampling, we can see that in both cases the model can describe the traffic patterns in most channels accurately, but there are channels in which the traffic pattern is only described accurately when using sampling. Figure \ref{fig:anomalies_CDF} shows the CDF of the ratio of the anomaly measures (Unknown+Miss+Retransmit) over all of the channels, with and without sampling. We can see from the figure that when using burst-based sampling there are much fewer channels with significant number of anomalies---the distribution ``tail'' is thinner.

\begin{table}[]
\centering
\caption{Burst Structure Accuracy}
\label{table:burst_structure_results}
\begin{tabular}{|l|l|l|}
\hline
\textbf{model\textbackslash{}accuracy ratio}                                  & \textbf{Bad beginning} & \textbf{\begin{tabular}[c]{@{}l@{}}Bad ending\end{tabular}} \\ \hline
\textbf{k-phase}                                                            & 1.03\%                 & 1.43\%                                                                           \\ \hline
\textbf{k-phase+sampling}                                                   & 0.013\%                 & 0.018\%                                                                          \\ \hline
\textbf{Burst-DFA}                                                      & 0.67\%                 & 0.53\%                                                                           \\ \hline
\textbf{\begin{tabular}[c]{@{}l@{}}Burst-DFA +sampling\end{tabular}} & 0.008\%                 & 0.006\%                                                                          \\ \hline
\end{tabular}
\end{table}

\subsection{Model Permissiveness}
\begin{figure}[t]
\centering
  \includegraphics[ scale = 0.4, trim = 40 20 20 20]{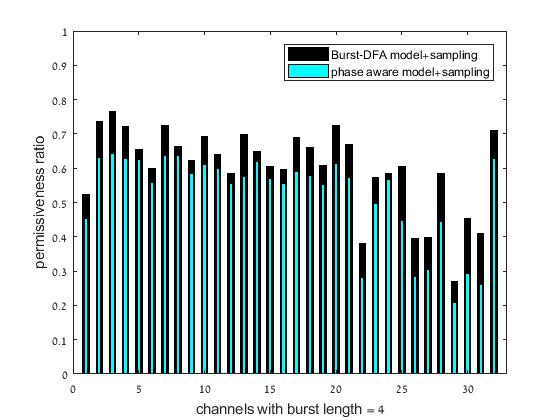}
   \hfill
\caption{Permissiveness ratio $R_{perm}$ for channels with average burst length of 4, with training set sampling}
    \label{fig:permissiveness_with_sampling}
\end{figure}

As mentioned, we expect the permissiveness of the k-phase model to be lower than the permissiveness of the Burst-DFA model, since our model has more restrictions on the traffic pattern. Figure \ref{fig:permissiveness_with_sampling} shows $R_{perm}$ for all of the channels with average burst length of 4, for the k-phase and burst-DFA models, with burst-based sampling. We can see that as expected, the level of permissiveness improves by 14\% on average compared to the Burst-DFA model. 

\section{Conclusions and future work}
In our research, we analyzed a large corpus of Modbus traffic recorded at a large scale water treatment plant in the U.S. Previous research on this data corpus suggested a DFA-based model to describe the traffic; however these models did not achieve the accuracy necessary to maintain a high-fidelity system with low false alarms. In this work we showed how to improve the model fidelity to the traffic in the network while maintaining low permissiveness (detecting anomalies). 
%that on average, the accuracy of the previous model deteriorates over time. We then showed 

To achieve this we showed how the network traffic has different phases over time. Based on this observation, we developed a method to describe the traffic patterns at different points in time using adjacency matrices, and introduced a measure of similarity between the traffic patterns. Then we developed a method to automatically assign the traffic into clusters based on the similarity metric introduced---i.e., an algorithm to detect the different traffic phases automatically. Next, we introduced a novel burst-based training set sampling method, which allows for training set assembly from the entire duration of the recording. The burst-based sampling method comes to ensure training on traffic from all traffic phases. We then developed a new k-phase model that incorporates the different traffic phases, by creating a unique DFA for each traffic phase detected. We also suggest a modified metric for the permissiveness of the model, that includes a method to count the unique number of walks across a set of directed graphs. The new k-phase model achieves a low False Alarm Rate, while limiting the permissiveness of the model. Finally, we showed that when using the new model, together with the burst-based sampling method, we can improve the accuracy and lower the permissiveness compared to previous models. We can achieve up to 98.9\%-99.99\% accuracy when using the burst-based sampling, and the k-phase model improves the permissiveness by approximately 14\%.

Furthermore, the automatic identification of phase changes in the traffic has value beyond anomaly detection. Labeling these phases can help the operator understand the different states of the controlled equipment, and has the potential to bridge the semantic gap that exists in anomaly detection models.

Future work includes testing our model on other large scale datasets, testing it on longer recordings, and also testing the model's performance during true attacks on the network. We are also interested in exploring the connection between the traffic phases and the actual tasks the controlled equipment is performing and the connection between the statistical characteristics of the phase shifts and the type and designation of the controlled equipment.

{\bf Acknowledgements}
This work was supported by a grant from the United States-Israel Binational Science Foundation (BSF), Jerusalem, Israel and the United States National Science Foundation (NSF) CNS-\#1718848. 
This material was also supported by a grant from the Interdisciplinary Cyber-Research Center at TAU.
%, and by the Laboratory for Analytic Sciences (LAS). Any opinions, findings, conclusions, or recommendations expressed in this material are those of the authors and do not necessarily reflect the views of the LAS and/or any agency or entity of the United States Government. 

\balance
\bibliographystyle{plain}
\bibliography{amitbib}  % amit.bib is the name of the Bibliography in this case
%Appendix A
%\section{Headings in Appendices}

\end{document}